\documentstyle[aps,twocolumn,psfig]{revtex}
\begin{document}
\title{The low-energy scale of the periodic Anderson model}
\author{Th.\ Pruschke$^a$, R.\ Bulla$^b$ and M.\ Jarrell$^c$}
\address{$^a$Institut f\"ur Theoretische Physik, Universit\"at Regensburg,
D-93040 Regensburg\\
$^b$Institut f\"ur Physik, Universit\"at Augsburg, D-86135 Augsburg\\
$^c$Department of Physics, University of Cincinnati, Cincinnati, OH
45221}
\draft
\maketitle
\begin{abstract}
Wilson's Numerical Renormalization Group method is used to study the
paramagnetic ground state of the periodic Anderson model within the
dynamical mean-field approach.  For the particle-hole symmetric model,
which is a Kondo insulator, we find that the lattice Kondo scale $T_0$
is strongly enhanced over the impurity scale $T_K$; $T_0/T_K\propto
\exp\{1/3I\}$, where $I$ is the Schrieffer-Wolff exchange coupling.  In
the metallic regime, where the conduction band
filling is reduced from one, we find characteristic signatures of 
Nozi\`eres exhaustion scenario, including a strongly reduced lattice Kondo 
scale, a significant suppression of the states available to screen 
the f-electron moment, and a Kondo resonance with a strongly enhanced 
height.  However, in contrast to the quantitative predictions of Nozi\`eres,
we find that the $T_0\propto T_K$ with a coefficient which depends
strongly on conduction band filling.
\end{abstract}
\section{Introduction\label{sec:1}}
Lanthanide or Actinide-based heavy fermion compounds\cite{steglich} can 
be viewed as a paradigm for strong correlation effects in solids. These 
systems comprise a great variety of different low-temperature states, 
including paramagnetic metals with either a Fermi liquid or non-Fermi 
liquid properties\cite{steglich,loehneysen}, long range magnetic order, 
superconductivity or coexistent magnetism and superconductivity\cite{steglich}. 
While it is very interesting and challenging to understand the physics of
these ordered ground states, not even the basic ingredient, namely the 
physics of the paramagnetic metal, has been fully captured yet on a
microscopic basis. The common understanding is that the physical properties 
of heavy fermion compounds are largely determined by spin-flip scattering 
between spins localized on the Lanthanide or Actinide sites and delocalized 
conduction quasi particles usually formed by the $d$-states in the system. In 
the case of dilute compounds this spin-flip scattering leads to the well-known
Kondo effect, i.e.\ the screening of the localized spins by the
conduction electrons and the formation of a Fermi liquid with a
low-energy scale set by the so-called Kondo temperature, $T_{\rm K}$ depending
exponentially and non-perturbatively on the system parameters
\cite{alex}. Interestingly,
the physics of the metallic phase of concentrated systems can in some
cases, e.g.\ CeAl$_3$, CeB$_6$ or CeCu$_6$ \cite{steglich}, be at least
qualitatively understood by a picture of independent, but coherent
Kondo scatterers, with the low-energy scale set by the Kondo
temperature of the dilute system. However, for UPt$_3$, URu$_2$Si$_2$
or Yb$_4$As$_3$ \cite{Yb4As3} or the compound LiV$_2$O$_4$ recently
characterized as heavy fermion system \cite{LiV2O4} there seem to
exist two distinct energy scales; one high temperature scale, $T_{\rm  K}$, 
describing conventional incoherent Kondo scattering, and a much
smaller scale, $T_0$, also called coherence scale in literature, which 
marks the onset of Fermi liquid formation characterized by quasi particles
with strongly enhanced effective masses.

This apparent discrepancy in the experimental situation has triggered numerous 
theoretical investigations, from which two dominant competing scenarios 
have emerged.

One scenario, based on a diagrammatic
perturbation theory \cite{kuramoto,grewe}, succeeded in approximately
mapping the concentrated system onto a set of independent, effective
impurities which at low temperature form a coherent Fermi liquid state
due to the lattice periodicity. The theory further predicts the
existence of a {\em{single}} energy scale which uniquely determines the
high energy properties like Kondo screening as well as the formation
of the low-temperature Fermi liquid. Most important, apart from
moderate renormalizations which usually lead to an enhancement, this
lattice scale is given in terms of the energy scale of the
corresponding diluted system \cite{grewe} and the one particle
properties close to the Fermi energy are well described by a picture
of hybridizing bands, leading to a density of states with a (pseudo-)
gap slightly above the Fermi energy \cite{cox}. The large effective masses can be 
accounted for by the observation that the Fermi energy lies in the
region with flat bands \cite{grewe,NCA2}. 
A further theoretical study, based on a variational treatment of the system 
with a Gutzwiller wave function and employing the Gutzwiller
approximation, seems to support this result in the
sense that the energy scale calculated is enhanced over the
corresponding scale of the dilute system, i.e.\ there is quite likely
only one relevant energy scale. However, this variational scale is
more strongly renormalized \cite{ueda}.

The other scenario is based on Nozi\'eres argument that in concentrated 
systems there will generally not be enough conduction states available 
to screen all of the localized spins.  
This situation is engendered by the fact that only the band states within 
$T_K$ of the Fermi surface can effectively participate in screening the 
local moments.  The number of screening electrons can be estimated as 
$\rho^c_0(0) T_K\ll1$, where $\rho_0^c(0)$ is the conduction band density of states 
at the Fermi level.
Thus, in a concentrated system one should encounter an intermediate   
temperature regime where all band states available for screening are 
``exhausted'' -- from which 
the name ``exhaustion physics'' was coined -- and the system starts
to resemble an incoherent metal where only part of the spins will be
screened \cite{nozieres}. However, since the Kondo screening clouds
are not pinned to a particular site they can move through the system 
with an effective hopping matrix element and a residual strong local
correlation, because there can never be two screening
clouds on the same site.
Based on these arguments Nozi\`eres thus
suggests, that at low temperatures the system effectively behaves
similar to a Hubbard model with a small number of holes and strong
local Coulomb repulsion.
The remaining entropy can now be quenched by 
forming a state with either long-range antiferromagnetic order or short 
ranged antiferromagnetic fluctuations, which give rise to a small low-energy
scale and a corresponding heavy Fermi liquid, too \cite{infdrev}.
It is important to note that these phenomenological arguments by
Nozi\`eres do not make any reference to a particular model or specific 
parameter regime of a model like the periodic Anderson model; thus, if 
Nozi\`eres arguments are correct, exhaustion physics should actually be the
generic situation.

Although both scenarios seem to be in accordance with some experimental 
facts, they apparently fail to capture the whole story. Moreover, in some 
cases they rely either on approximations that are difficult to control on a 
microscopic basis or are entirely based on phenomenological arguments, as it 
is the case with Nozi\`eres' point of view. In order to discriminate which 
ansatz is correct or outline the borders where each of the two scenarios may 
be valid, a more thorough microscopic study of the properties of the paramagnetic
phase is highly desirable.

Recently, 
Quantum Monte Carlo (QMC) simulations of the periodic Anderson model 
(PAM) within the Dynamical Mean-Field Theory (DMFT) \cite{DMFT,infdrev,Geo96} 
have shown that when the conduction band is nearly half
filled\cite{PAMinfd2}, there is a single scale consistent with the
predictions of Rice and Ueda \cite{ueda},
but obviously inconsistent
with Nozi\`eres' picture of exhaustion. On the other hand, when the
conduction band filling is significantly reduced, the states available
for screening of the local moments appear to be depleted near the
Fermi surface and the lattice coherence scale is strongly reduced from
the corresponding impurity scale \cite{niki1,niki3}.  A protracted
evolution of the photoemission spectra \cite{niki2}, and transport
\cite{niki3} are predicted and can be understood in terms of a
crossover between the two scales. Although a quantitative relationship
between the two scales could not be established, the reduction was
ascribed to exhaustion \cite{niki2}. Nozi\`eres subsequently argued
that exhaustion should lead to a significant reduction of the lattice
scale and predicted that $T_0$ is {\em{at most}} $ N_d(0)T_K^2$
\cite{nozieres2}.

In this contribution we present results for the zero-temperature properties 
of the periodic Anderson model obtained within the DMFT in conjunction with 
Wilson's numerical renormalization group (NRG) calculations
\cite{NRG}. We will show that for systems with approximately half
filled conduction band, one finds a single energy scale, in accordance
with the independent impurity picture and which qualitatively behaves
as predicted by Rice \& Ueda \cite{ueda};
moreover, in this parameter regime we do not find any hint towards the
occurrence of exhaustion physics; which raises the question to what
extent Nozi\`eres' phenomenological argument, that only a fraction
$\rho_0^c(0)T_{\rm K}\ll1$ of the band states can contribute to the
screening of the f-spins, is valid.

In order to observe a behavior at all that is in at least qualitative
agreement with Nozi\`eres' exhaustion scenario, one actually
has to strongly reduce the carrier concentration of the conduction
band. For these low-carrier systems the physical behavior completely
changes and can now be understood in the framework of
exhaustion. However, a quantitative comparison of the lattice energy
scale and the one of the corresponding impurity model is possible with
our method, and indicates a strikingly different relation between the
two than recently predicted by Nozi\`eres \cite{nozieres2} even in
this parameter regime.

The paper is organized as follows. In the next section we will
introduce the periodic Anderson model and its treatment within the
DMFT using NRG. In section \ref{sec:3} we will discuss recent
developments in understanding the physics of the PAM in some detail
and provide a basis to interpret the NRG results presented in section
\ref{sec:4}. The paper will be concluded by a summary and outlook in
section \ref{sec:5}.
\section{Model and Method\label{sec:2}}
\subsection{Periodic Anderson Model}
The standard model used to describe the physics of heavy fermion
systems is the periodic Anderson model (PAM)
\begin{equation}\label{equ:1}
\begin{array}{ll}
\displaystyle
H=&
\displaystyle\sum_{{\bf k}\sigma}\varepsilon_{\bf k}c^\dagger_{{\bf
      k}\sigma}c^{\phantom{\dagger}}_{{\bf k}\sigma} + \varepsilon_f
  \sum_{{\bf k}}f^\dagger_{{\bf k}\sigma}f^{\phantom{\dagger}}_{{\bf
        k}\sigma}+
      U\sum\limits_in^f_{i\uparrow}n^f_{i\downarrow}\\[5mm]
&\displaystyle + \sum_{{\bf k}\sigma}V_{\bf k}\left(c^\dagger_{{\bf
            k}\sigma}f^{\phantom{\dagger}}_{{\bf k}\sigma} +
        \mbox{h.c.}\right) \;\;.
\end{array}
\end{equation}
In (\ref{equ:1}), $c^{\dagger}_{{\bf k}\sigma}$ creates a conduction
quasi particle with momentum ${\bf k}$, spin $\sigma$ and dispersion
$\varepsilon_{\bf k}$, with $\frac{1}{N}\sum\limits_{\bf
  k}\varepsilon_{\bf k}=\varepsilon_c$ as its center of mass;
$f^\dagger_{{\bf k}\sigma}$ creates an
f-electron with momentum  ${\bf k}$, spin $\sigma$ and energy
$\varepsilon_f$ and $n^f_{i\sigma}$ is the number operator for
f-electrons at lattice site $i$ with spin $\sigma$. The localized
f-states experience a Coulomb repulsion $U$ when occupied by two
electrons and the two subsystems are coupled via a hybridization
$V_{\bf k}$. Although it is in general a crude approximation we will
assume for computational reasons $V_{\bf k}=\mbox{const.}$ for the
rest of the paper.

Given the usually rather complex crystal structure of heavy fermion
compounds, the question to what extent the simple model (\ref{equ:1}), 
which describes orbitally non-degenerate f-states, is
appropriate. However, especially in Ce-based systems the relevant
configuration is f$^1$, whose multiplicity will in general be reduced
to a Kramers doublet in the crystal field; the other crystal field
multiplets are generally well separated from this ground state doublet 
\cite{steglich}. The situation is more complicated in Uranium
compounds, where with equal probability a f$^2$ state can be the lowest
configuration. In this case the assumption of a Kramers doublet
is of course not justified. Nevertheless, even for these compounds the 
PAM seems to provide an at least qualitatively correct description, so 
that we assume it to be the relevant model for those compounds, too.

In contrast to e.g.\ the Hubbard model, where for one dimension an
exact solution via Bethe ansatz is available and a combination of
many-body techniques was successful to exploit the ground state
properties exactly, no such exact benchmark is available for the
PAM. The only limit where the physical properties are almost
completely known is the impurity version of the model (\ref{equ:1})
(single impurity Anderson model, SIAM) \cite{alex}, where a single site with
f-states exists which couples to the conduction states. In this limit
one finds for $|\varepsilon_f|/V,\;(\varepsilon_f+U)/V\gg1$ the
mentioned Kondo behavior, namely a screening of the local moment by
the conduction states with an energy scale  (Kondo temperature)
\begin{equation}\label{equ:2}
\begin{array}{l@{\;=\;}l}
\displaystyle
T_{\rm K} & \displaystyle\lambda\sqrt{I} \exp\left(-\frac{1}{I}\right)\\[5mm]
\displaystyle I & \displaystyle2V^2\rho_0^c(0)\left|\frac{1}{\varepsilon_f}-\frac{1}{\varepsilon_f+U}\right|
\end{array}
\end{equation}
with $\rho_0^c(0)$
the density of states at the Fermi energy of the conduction band and
$\lambda$ a cut-off energy. Note, that $T_{\rm K}$ depends exponentially 
on the system parameters, and especially non-analytically on $V^2$.

It is important to stress that the occurrence of the Kondo temperature
is intimately connected to the thermodynamic limit, where a (quasi-)
continuum of band states exists at the Fermi energy.  This 
property makes reliable calculations with techniques usually suitable
to treat finite-sized two or three dimensional systems, like exact
diagonalization or quantum Monte Carlo, of limited value, since
they can only handle small to moderately sized systems. For the
quantum Monte Carlo an additional problem is the sign problem, that
becomes increasingly serious as $U$ and the system size increase. Direct 
perturbational approaches like e.g.\ FLEX are by construction restricted to small
values of $U$ and generally fail to capture even the basics of the
Kondo physics. Notable exceptions occur again for the SIAM. Here,
straight forward second order perturbation theory in $U$ ({\em not self-consistent}) is able to
reproduce at least qualitative features in the special case of
full particle-hole symmetry \cite{yamada,zlatic}. More recently, a
novel ansatz using a perturbation theory about the unrestricted
Hartee-Fock solution for the SIAM \cite{glossop} has been shown to give
even quantitatively correct results for spectral functions
and energy scales of the SIAM in the Kondo limit \cite{BGLP}.
\subsection{Dynamical Mean-Field Theory}
The other non-trivial limit, where an exact solution of the PAM (\ref{equ:1}) 
becomes at least in principle possible is the limit of large
dimensionality, where the dynamical mean-field theory (DMFT) becomes exact
\cite{DMFT,infdrev,Geo96}. Here, the renormalizations due to the local
Coulomb repulsion become purely local \cite{DMFT}, i.e.\ one obtains a 
one-particle self energy independent of momentum
\begin{equation}\label{equ:4}
\Sigma^f({\bf k},z)\to \Sigma^f(z)\;\;.
\end{equation}
This property may be 
used to remap the lattice problem onto an effective SIAM again
\cite{brandt,QMC_MJ}. The
non-trivial aspect of the theory comes about by the fact that the
medium coupling to the effective impurity is a priori not known but
has to be determined self-consistently in the course of the
calculation \cite{self_consistency}. In particular, the
self-consistency relation for the PAM reads
\begin{equation}\label{equ:3}
G^{\rm loc}(z)\begin{array}[t]{@{\;=\;}l}\displaystyle\int
d\epsilon\frac{\rho_0^c(\epsilon)}{z-\varepsilon_f-\Sigma^f(z)-\frac{V^2}{z-\epsilon-\varepsilon_c}}\\[5mm]
\displaystyle\frac{1}{z-\varepsilon_f-\tilde{\Delta}(z)-\Sigma^f(z)}\stackrel{!}{=}G^{\rm 
  SIAM}(z)\;\;,
\end{array}
\end{equation}
with $\rho_0^c(\epsilon)$ the density of states of the bare conduction 
band in (\ref{equ:1}). The second line in (\ref{equ:3}) defines a
generalized hybridization function $\tilde{\Delta}(z)$, which
implicitly depends on $\Sigma^f(z)$ and is thus modified from its
non-interacting form by the presence of correlated f-electrons on other
sites in an averaged way.

The remaining task is to solve the effective SIAM defined by the set of 
parameters $\{\varepsilon_f,\; U\}$ and the generalized
Anderson width 
\begin{equation}\label{equ:6}
\Gamma(\omega)=-\Im m\{\tilde{\Delta}(\omega+i\eta)\}\;\;.
\end{equation}
Note that the self-consistency condition
(\ref{equ:3}) requires the knowledge of a dynamical quantity, viz the
one-particle Green function for all frequencies. This immediately
rules out techniques like Bethe ansatz, since it is impossible to
calculate dynamical correlation functions with this method. To perform
this task nevertheless, a variety of different methods
have been developed and applied during the past decade: Quantum Monte Carlo
simulations \cite{QMC_MJ}, exact diagonalization \cite{Geo96}, extended
second order perturbation theory (iterated perturbation theory, 
IPT) \cite{Geo96,IPT}, resolvent perturbation
techniques \cite{NCA2,NCA}, local-moment approach \cite{glossop}
and Wilson's NRG \cite{Sak94,Sak95,BPH}.

In this contribution we used the last method, for the following
reasons. First, it is tailored to capture the low-energy physics of
the Kondo problem with high accuracy. Second, it is able to identify
exponentially small energy scales. Third, it is non-perturbatively and 
thus also produces the correct dependence of the Kondo temperature on
the parameters (see (\ref{equ:2})) and, last but not least, it has only
little numerical restrictions on the choice of
model parameters. Together with past years developments
\cite{BPH,SCES99} we are able to study the low-temperature properties
of the PAM with the NRG with good accuracy.
\subsection{Details of Wilson's NRG}
Quite generally, the NRG is based on a logarithmic discretization of the
energy axis, i.e.\ one introduces a parameter $\Lambda>1$ and divides 
the energy axis into intervals $[\Lambda^{-(n+1)},\Lambda^{-n}]$ for
$n=0,1,\ldots,\infty$ \cite{alex,NRG}. With some further manipulations \cite{BPH} one can
map the original model onto a semi-infinite chain, which can be solved 
iteratively by starting from the impurity and successively adding
chain sites. Since the coupling between two adjacent sites $n$ and $n+1$ vanishes  like $\Lambda^{-n/2}$ for large $n$, the low-energy
states of the chain with $n+1$ sites are determined
by a comparatively small number $N_{\rm states}$ of states close to the
ground state of the $n$-site system. In practice, one
retains only those $N_{\rm states}$ from the $n$-site chain to set up the
Hilbert space for $n+1$ sites and thus prevents the usual exponential
growth of the Hilbert space as $n$ increases.
Eventually, after $n_{\rm NRG}$ sites have been included in the
calculation, adding another site will not change the spectrum significantly
and one terminates the calculation.

It is obvious, that for any $\Lambda>1$ the NRG constitutes an approximation
to the system with a continuum of band states but becomes exact in the
limit $\Lambda\to 1$. Performing this limit is, of course, not
possible as one has to simultaneously increase the number
of retained states to infinity. One can, however, study the 
$\Lambda$- and $N_{\rm states}$-dependence of the NRG results and perform the 
limit $\Lambda\to 1$, $N_{\rm states}\to\infty$ by extrapolating these
data. Surprisingly one finds that the dependence of the NRG results on 
$\Lambda$ as well as on the cut-off $N_{\rm states}$ is extremely mild, in 
most cases is a choice of $\Lambda=2$ and $N_{\rm states}=300\ldots500$
sufficient \cite{BGLP}.

While the knowledge of the states is sufficient to calculate
thermodynamic properties, the self-consistency (\ref{equ:3}) requires
the knowledge of the one-particle Green function or, equivalently the
knowledge of the single-particle self energy $\Sigma^f(z)$. Since the
NRG scheme works with a discretization of the energy axis, it
corresponds to a discrete system and by construction the Green
function consists of a set of poles and an appropriate coarse-graining 
procedure has to be applied. During the last 15 years
considerable progress has been made to extract dynamical properties
with the NRG, too, and it has been shown to give very accurate
results also for e.g.\ dynamical one- and two-particle and also
transport properties \cite{Sak89,Cos94}. For dynamical properties the
NRG works best at $T\!=\!0$, and various dynamic correlation functions
can be calculated with an accuracy of a few percent. Although less well defined
for finite temperatures, its extension to $T>0$ also shows very good
agreement with exact results \cite{Sak95,Cos94}. It is quite remarkable
as no sum-rules (Friedel sum rule, total spectral weight etc.) must be used as
input for these calculations. On the contrary, they can serve as an independent
a posteriori check on the quality of the results. The first application
of the NRG to the DMFT known to us is the work of Sakai et al.\
\cite{Sak94,Sak95} where the symmetric Hubbard and periodic Anderson
model in the metallic regime have been studied. In their papers, the
authors point out some difficulties in the progress of iterating the
NRG results with the DMFT equations, which are in our opinion largely
related to the necessary broadening of the NRG spectra.

As has been shown in the work of Bulla et al.\ \cite{BPH}, these
difficulties can be circumvented if, instead of
calculating $G^f(z)$ and extracting $\Sigma^f(z)$ from it, one
calculates the self energy directly via the relation
\begin{equation}\label{equ:5}
\begin{array}{l@{\:=\:}l}
\displaystyle\Sigma^f_\sigma(z) & \displaystyle U \frac{F^f_\sigma(z)}{G^f_\sigma(z)}\\[5mm]
\displaystyle F^f_\sigma(z) & \langle\langle f_{\sigma} f^\dagger_{\bar{\sigma}}
f_{\bar{\sigma}},f^\dagger_{\sigma} \rangle\rangle_z\;\;,
\end{array}
\end{equation}
which originates from the equation of motion
\begin{equation}\label{equ:5a}
\begin{array}{l@{\:=\:}l}
\displaystyle zG^f_\sigma(z) &
\displaystyle 1+\langle\langle[f^{\phantom{\dagger}}_\sigma,H_{\rm
  SIAM}],f^\dagger_\sigma\rangle\rangle(z)\\[5mm]
 &
\displaystyle
1+\left(\varepsilon_f+\Delta(z)\right)G^f_\sigma(z)+UF^f_\sigma(z)
\end{array}
\end{equation}
with
$$
\Delta(z)=\frac{1}{N}\sum\limits_{\bf k}\frac{|V_{\bf
    k}|^2}{z-\varepsilon_{\bf k}}\;\;.
$$
Both correlation functions $G_\sigma^f(z)$ and $F_\sigma^f(z)$ appearing in (\ref{equ:5}) can be
calculated with the NRG and it turns out that the quotient of them
gives a much better result for $\Sigma^f(z)$ than the use of $G^f(z)$ alone 
\cite{BPH}.

Let us note one particular problem in dealing with the PAM in the
DMFT+NRG. As we will see in the beginning of section \ref{sec:4}, the
effective hybridization for the PAM in the particle-hole symmetric
limit exhibits a pole right at the Fermi level (see Fig.\ \ref{fig:2}
on page \pageref{fig:2}). It is clear that such a pole will lead to
numerical difficulties. However, the NRG allows to deal with such a
structure in a very efficient way, namely by including this single
state represented by the pole into the definition of the ``impurity''
defining the beginning of the NRG chain.
\section{Recent Results}\label{sec:3}
Early studies of the PAM using the DMFT concentrated on the
particle-hole symmetric case $\varepsilon_c=0$, $2\varepsilon_f+U=0$,
i.e.\ $n_f=1$ and $n_c=1$
\cite{PAMinfd2,Sak95,PAMinfd1}. To solve the SIAM, the authors employed QMC
\cite{QMC_MJ,QMC} and NRG \cite{Sak95}. Although for this particular situation the
system is a band insulator for symmetry reasons, one can extract a
Wilson Kondo scale from the SIAM impurity susceptibility. Interestingly,
this energy scale is enhanced with respect to the corresponding SIAM
Kondo temperature in accordance with the results by Rice \& Ueda
\cite{ueda,PAMinfd2}. In addition, at low temperatures the system can order
antiferromagnetically \cite{PAMinfd2}. To avoid confusion and make
the discussion more transparent we will use $T_0$ and $T_{\rm K}$ from now on the
distinguish the relevant lattice energy scale and the Kondo scale, respectively.

One possibility to break the particle-hole symmetry is by depleting
the band filling via changing $\varepsilon_c$, but keeping $n_f\approx1$, 
which has been done with QMC for $U/V^2\approx4$ \cite{niki1,niki3,niki2}. 
The resulting phase diagram turns out to be quite interesting for several 
reasons. First, one finds a suppression of the antiferromagnetic order with 
decreasing $n_c$ and around $n_c=0.5$ a region with ferromagnetic order  
emerges \cite{niki1}, which for $n_c < 0.5$ can be 
attributed to a ferromagnetic effective RKKY-type exchange. Second,
from studies of the resistivity and optical conductivity
\cite{niki3,niki2}, one can infer that for $n_c\alt0.8$ the physics of
the paramagnetic metallic phase drastically change. While for
$n_c\agt0.8$ there seem to exist only one energy scale $T_0\agt T_{\rm 
  K}$, the resistivity data suggest, that for $n_c\alt0.8$ the onset
of coherence is marked by a temperature scale $T_0\ll T_{\rm K}$
\cite{niki3,niki2}. In addition, the spectral function drastically
changes as one decreases $n_c$ and the effective Anderson width
(\ref{equ:6}), which for $n_c\approx1$ has a peak at the Fermi energy, 
starts to develop a dip for $\omega=0$. These observations were taken
as fingerprints of Nozi\'eres exhaustion scenario \cite{niki3,niki2}.
Further evidence for this interpretation comes from the work
by Vidhyadhiraja et al.\ \cite{krish99}, which is based on $2^{nd}$
order perturbation theory in $U$ (IPT). These authors, too, find a
similar behaviour as function of $n_c$. In addition, since their
method allows to study $T=0$, they could extract the energy scale
$T_0$ from their data. Interestingly, they found a relation between
$T_0$ and $T_{\rm K}$ of the form $T_0\propto (T_{\rm K})^2$ for their 
results, which is precisley the behaviour predicted by the
phenomenological theory of Nozi\'eres \cite{nozieres2}. However, we note that
this particular relationship $T_0 \propto T_K^2$ in \cite{krish99} was found
only if the width of the Kondo peak was used to estimate $T_0$. When the ratio
of the masses was used instead, the relationship rather follows $T_0 \propto T_K$ \cite{georges2}.

The estimate of $T_0$ coming from Gutzwiller ansatz calculations on the other
hand apparently fail to capture the essential physics in this parameter regime,
since they predict a ratio $T_0/T_{\rm K}>1$ for all values of $n_c$ \cite{ueda}.

Although the results from QMC and IPT are at a first glance very convincing, there
remain some problems. First, both series of calculations were done
with a comparatively small value of $U/V^2\approx4$ and a systematic study of
the behaviour as function of $U/V^2$ especially for larger values is
clearly necessary. Second, the results were either obtained by QMC
\cite{niki3,niki2} or IPT \cite{krish99}. However, for large $U/V^2$,
the identification of exponentially small energy scales with QMC is
problematic due to its restriction to finite temperatures. The IPT on the other
hand leads to ambiguous results as mentioned before; in addition, as
a perturbational approach in $U$, it certainly cannot
produce exponentially small energy scales. Thus, for a quantitative
description of the low-temperature phase and especially a reliable
calculation of the low-energy scale $T_0$, a non-perturbative
technique at $T=0$ is necessary.
\section{NRG results}\label{sec:4}
Such a method has become available recently by the application of
Wilson's NRG \cite{NRG} to the DMFT
\cite{Sak94,Sak95,BPH}, which we use to study the paramagnetic phase of the
PAM within the DMFT at $T=0$. In order to perform the energy integral in 
(\ref{equ:3}) analytically and to be able to make contact to earlier results we
study a simple hypercubic lattice in the limit of dimensionality
$d\to\infty$. With the proper scaling \cite{infdrev} this leads to
$\rho_0^c(\epsilon)=\exp\left(-(\epsilon^2/t^\ast)^2\right)/(t^\ast\sqrt{\pi})$
for the noninteracting DOS of the band states. In the following we use 
$t^\ast=1$ as our energy unit.

\begin{figure}
\begin{center}\mbox{}
\psfig{figure=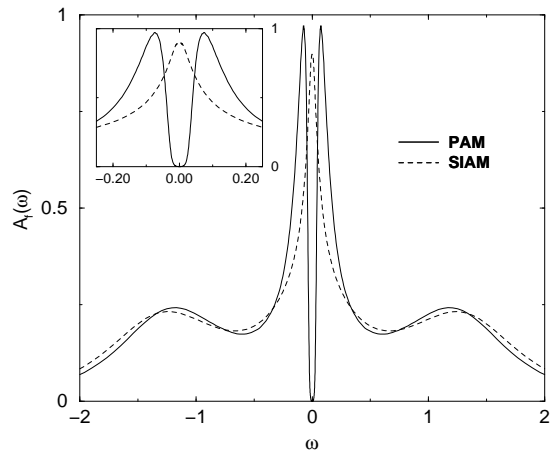,angle=-90,width=0.4\textwidth}
\end{center}
\caption[]{$f$-density of states $A_f(\omega)$ for SIAM (dashed line) and
PAM (full line) in the particle-hole symmetric case
$2\varepsilon_f+U=0$ and $\varepsilon_c=0$. The model parameters are
$U=2$ and $V^2=0.2$. The inset shows a blowup of the region around the 
Fermi level.}
\label{fig:1}
\end{figure}

\begin{figure}
\begin{center}\mbox{}
\psfig{figure=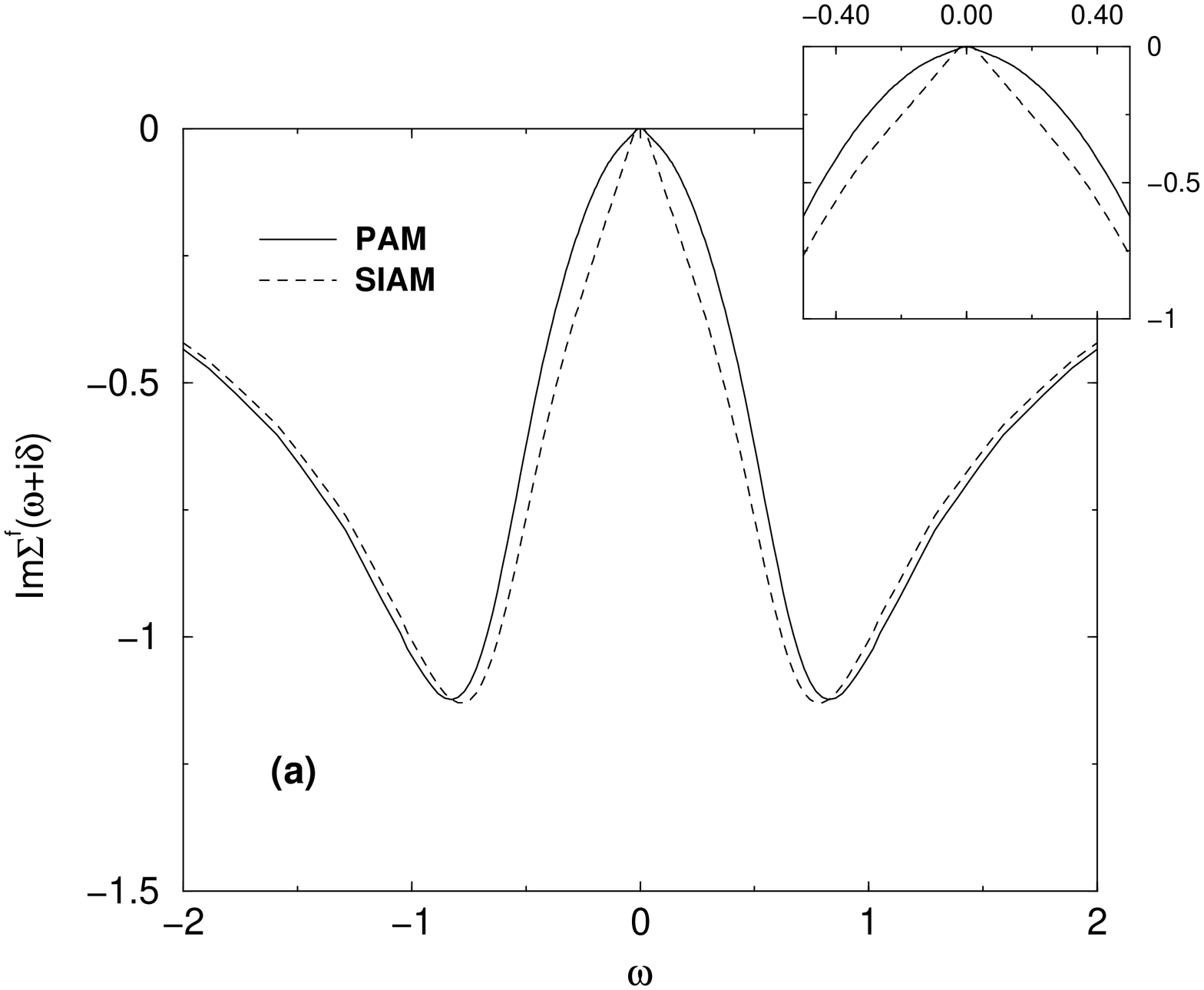,angle=-90,width=0.4\textwidth}\\
\mbox{}
\psfig{figure=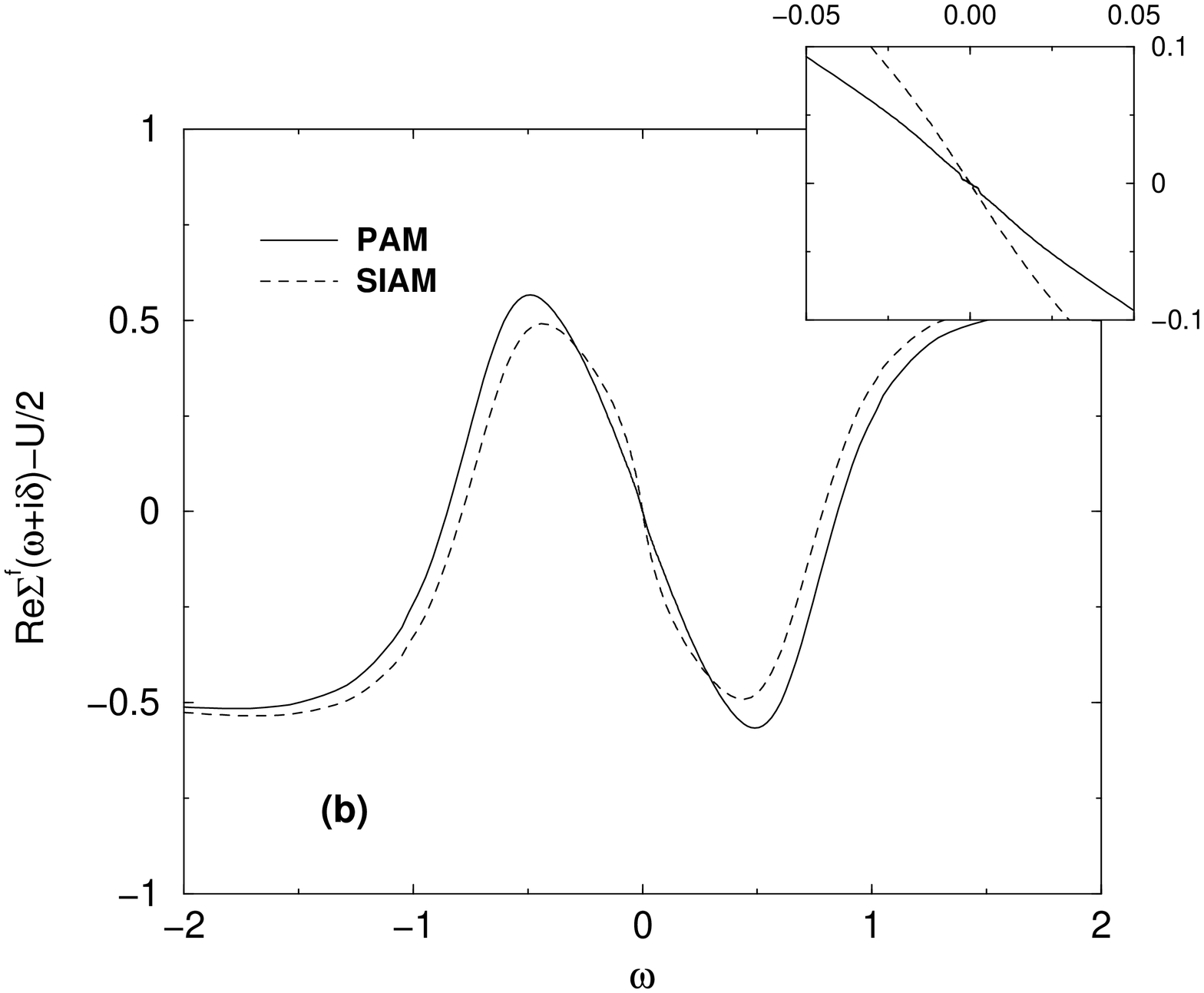,angle=-90,width=0.4\textwidth}
\end{center}
\caption[]{$\Sigma^f(\omega+i\delta)$ SIAM (dashed line) and
PAM (full line) in the particle-hole symmetric case and parameters
according to Fig.\ \ref{fig:1}. From the real part (b) the Hartree
term $U/2$ has been subtracted. The insets show a blowup of the region around the 
Fermi level.\label{fig:1a}}
\end{figure}

Let us start by briefly discussing the particle-hole symmetric
situation, i.e.\ $2\varepsilon_f+U=0$ and $\varepsilon_c=0$. Here, it
is expected from symmetry and shown by calculations
\cite{PAMinfd2,Sak95,PAMinfd1}, that the concentrated system exhibits a
hybridization gap. This is shown in Fig.\ \ref{fig:1}, where the
$f$-DOS $A_f(\omega)=-\frac{1}{\pi}\Im m\left\{G^f(\omega+i\delta)\right\}$
is plotted for the SIAM (dashed line) and the PAM (full line). The
model parameters are $U=2$ (i.e.\ $\varepsilon_f=-1$) and
$V^2=0.2$. The result for the SIAM nicely shows the
well-known structures, namely the charge excitation peaks at
$\omega\approx\pm U/2$ and \hfill the \hfill Abrikosov-Suhl \hfill
resonance \hfill (ASR) \hfill at \hfill the \hfill
Fermi
\noindent level. This latter structure, which can be regarded as an
effective local level, basically leads to the hybridization gap in 
the PAM. The enlarged view of the region around the Fermi energy
also shows, that the width of this hybridization gap and the width 
of the ASR are of similar order of magnitude. As we will dicuss in a
moment the corresponding energy scale in the lattice is actually
enhanced over $T_{\rm K}$, which sets the width of the ASR. This
result can readily be anticipated from the self-energy of the
$f$-states in Fig.\
\ref{fig:1a}. In the SIAM (dashed line) and the PAM (full line) \hfill 
one \hfill 
observes \hfill  a \hfill nice \hfill parabolic \hfill extremum
\noindent
in $\Im m\Sigma^f(\omega+i\delta)$
at the Fermi energy (see Fig.\ \ref{fig:1a}a), which is accompanied by
a linear region in $\Re e\Sigma^f(\omega+i\delta)$ in Fig.\
\ref{fig:1a}b; the slope is negative in both cases and definitely smaller for the PAM, as can be seen from the inset to
Fig.\ \ref{fig:1a}b. This means that, at least as far as the self
energy is concerned, the system can be viewed as a Fermi liquid with
a quasi-particle weight 
\begin{equation}\label{equ:7}
Z^{-1}=1-\left.\frac{d\Re
    e\Sigma^f(\omega)}{d\omega}\right|_{\omega=0}\equiv m^\ast/m\;\;,
\end{equation}
where we introduced the notion of the effective mass $m^\ast$. Note
that the above result, viz that the system is a Fermi liquid, is not a 
priori apparent from Fig.\ \ref{fig:1}, since the spectrum represents
an insulator. From the self-energy it is however evident that this
insulating state can be trivially accounted for by using the picture
of hybridizing quasi-particle bands, one of which is located at the
Fermi energy and describes an effective f-level.
The characterization of the particle-hole symmetric system as band
insulator is a priori not the only possible scenario. Indeed, for
small $V^2$ and $U\to\infty$ an alternative route to an insulating
state is via a Mott-Hubbard transition as found by Held et al.\ for
non-constant $V_{\bf k}$ \cite{held}. We did not observe such a
transition for the case $V_{\bf k}=$const.\ studied here for
values of $U$ as large as $U=10$, but also cannot exclude such a
scenario on the basis of the data available.

\begin{figure}
\begin{center}\mbox{}
\psfig{figure=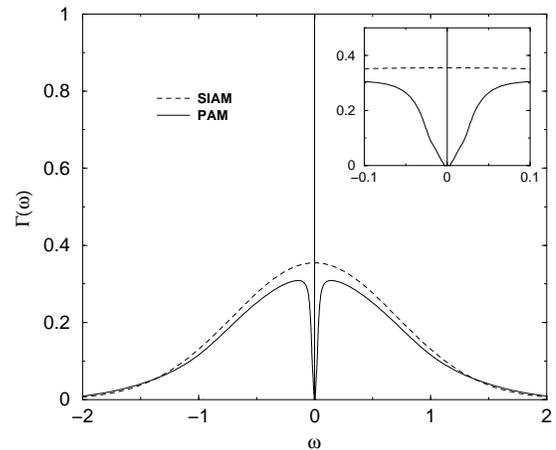,angle=-90,width=0.4\textwidth}
\end{center}
\caption[]{Anderson width $\Gamma(\omega)$ for SIAM (dashed line) and
PAM (full line) in the particle-hole symmetric case and parameters
according to Fig.\ \ref{fig:1}. The inset shows a blowup of the region around the 
Fermi level.\label{fig:2}}
\end{figure}

It is also quite illustrating to have a look at the effective Anderson 
width as defined by (\ref{equ:6}). This function is rather featureless 
for the SIAM (dashed line in Fig.\ \ref{fig:2}), but exhibits a very
pronounced structure near the Fermi
\noindent level for the PAM (full line in
Fig.\ \ref{fig:2}), namely a pseudo gap plus a strong peak right at
$\omega=0$; one can in fact show that the latter is a $\delta$-peak.
This $\delta$-peak can be understood as emerging from the sharp
quasi-particle band with dominantly f-character at the Fermi level.

\begin{figure}
\begin{center}\mbox{}
\psfig{figure=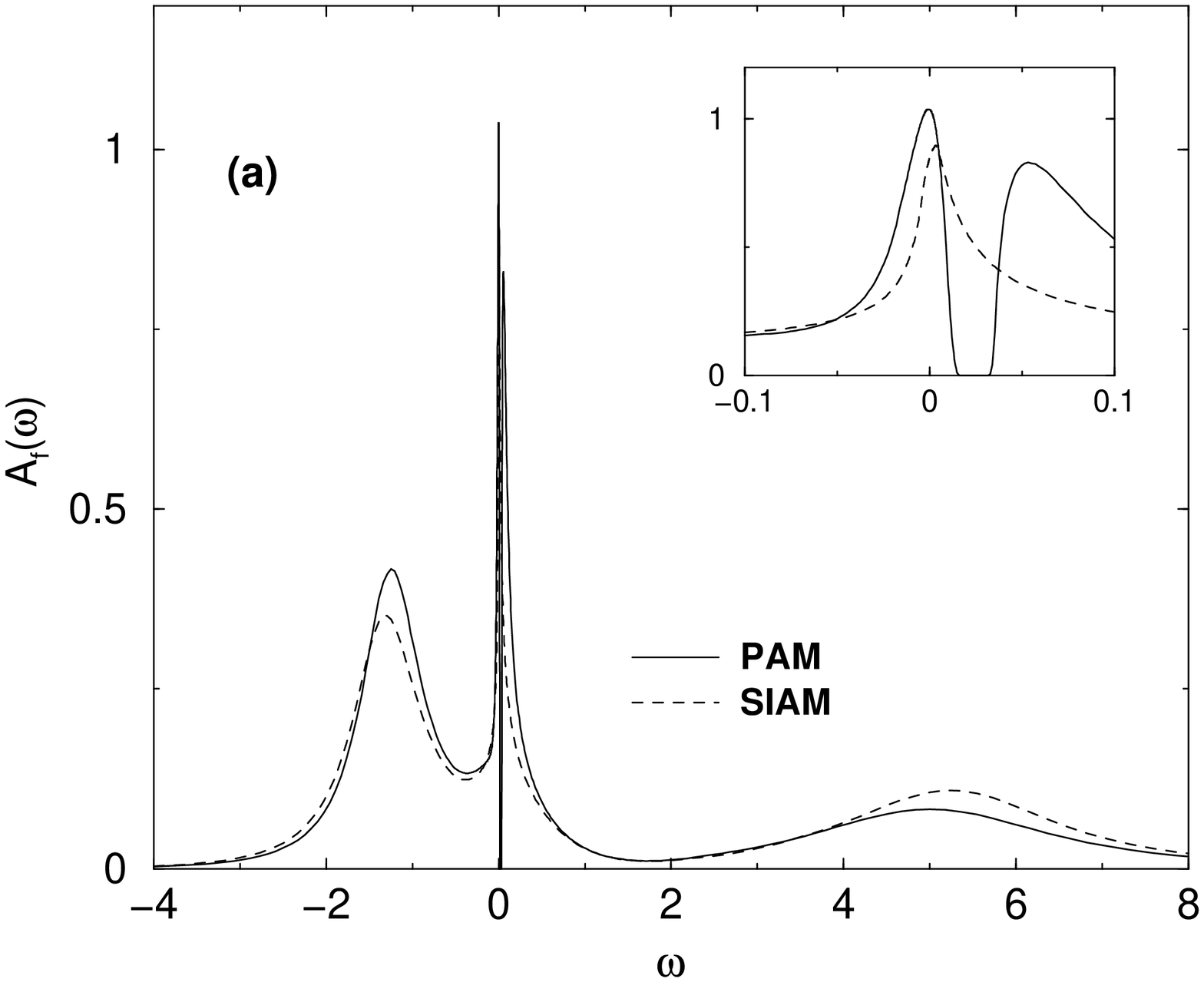,angle=-90,width=0.4\textwidth}\\
\mbox{}
\psfig{figure=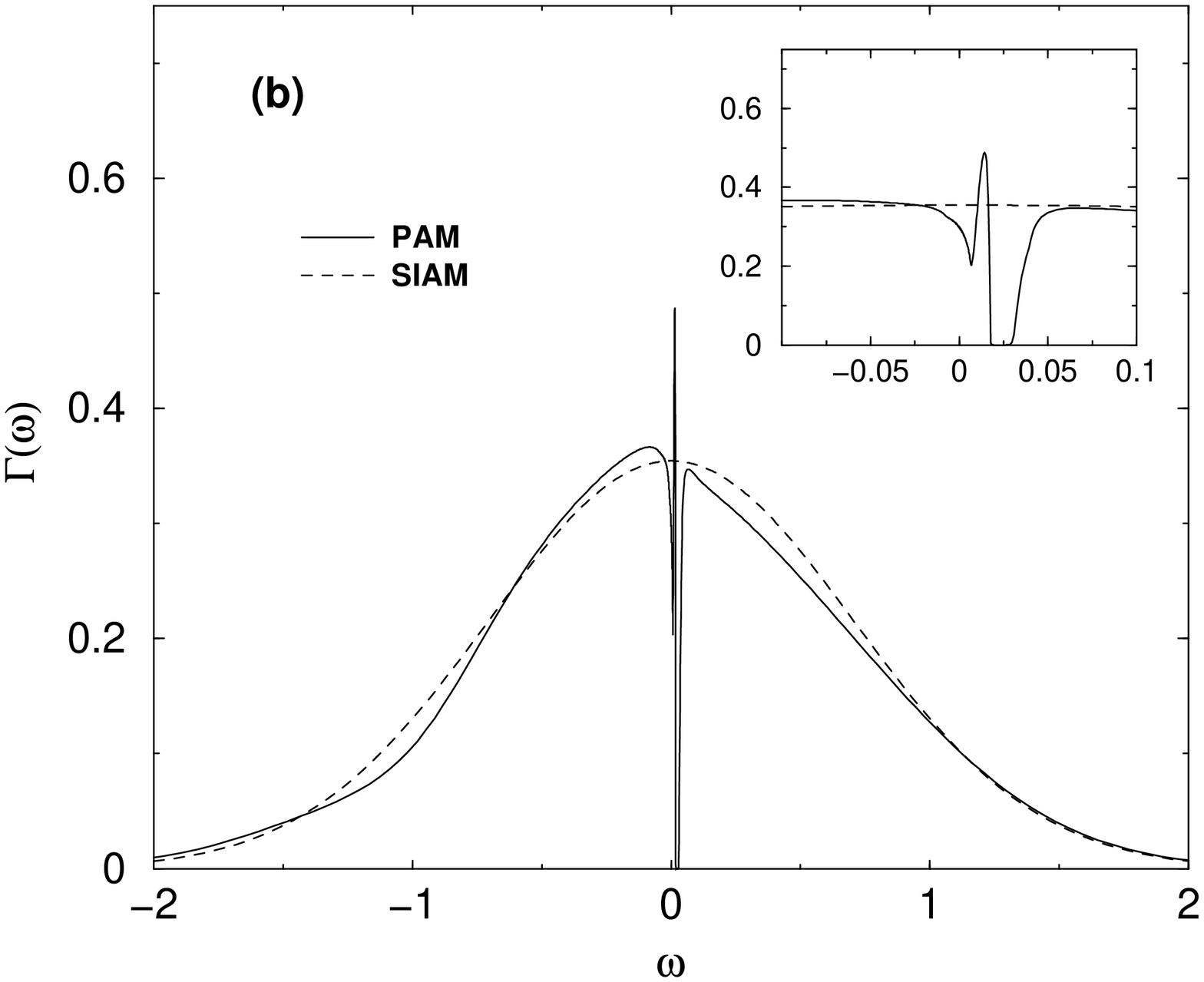,angle=-90,width=0.4\textwidth}
\end{center}
\caption[]{DOS $A_f(\omega)$ (a) and Anderson width $\Gamma(\omega)$
  (b) for SIAM (dashed line) and
PAM (full line) in the asymmetric case. The parameters are
$\varepsilon_c=0$, $\varepsilon_f=-1$, $U=6$ and $V^2=0.2$. The insets show a blowup of the region around the 
Fermi level.}
\label{fig:3}
\end{figure}

The results by QMC and IPT suggest that one can expect
drastic changes in the physics of the model if one breaks the
particle-hole symmetry. There are actually two possible routes to
accomplish this goal; the first is to keep
$\varepsilon_c=0$, and thus $n_c\approx1$, but increase $U$ so that $2\varepsilon_f+U>0$. An
example for the spectrum and the corresponding hybridization function
for $\varepsilon_f=-1$, $U=6$ and $V^2=0.2$ is shown in Fig.\ \ref{fig:3},
where we plot the DOS $A_f(\omega)$ (Fig.\ \ref{fig:3}a) and the
hybridization function $\Gamma(\omega)$ (Fig.\ \ref{fig:3}b). For
these particular parameters the occupancies are $n_c\approx1$
and $n_f\approx0.92$.
As in the particle-hole symmetric case one finds the characteristic
three-peak structure, again with a hybridization gap in the DOS of the 
lattice \sloppy\mbox{(full line, see inset to Fig.\ \ref{fig:3}a). Note that this
gap now is }
\noindent located above the Fermi energy and its width is visibly
larger than the width of the ASR in the SIAM; pointing again towards
an enhanced energy scale for the lattice, which is also confirmed by
an inspection of the self energy. A replica of this (pseudo-)
gap is also visible in the effective hybridization function of the
lattice in Fig.\ \ref{fig:3}b, which in addition shows a pronounced
peak slightly above the Fermi energy.
The origin of this peak is the same as for $n_f=1$; only it now has to 
be damped due to the finite lifetime  of the quasi-particles for
$\omega>0$.
It is quite interesting to note
that the value of $\Gamma(0)$ is actually reduced from its value in
the SIAM, but the average of  $\Gamma(\omega)$ over a region of the
order $T_0$ around the Fermi energy is enhanced. If one assumes that
it will be such an averaged value that determines the low-energy scale, 
one can readily understand that $T_0$ is enhanced over $T_{\rm K}$ by
this simple picture. We should mention that our results are very
similar to those obtained more than 10 years ago with resolvent
perturbation theories \cite{kuramoto,grewe,NCA2}. In fact, the
physical situation studied then was basically the same, namely a
particle-hole symmetric conduction band hybridizing with an asymmetric
$f$-level, however with $U=\infty$. The interpretation of the
structures in  $\Gamma(\omega)$, which may be viewed as the effective
DOS of the medium visible to the impurity, then leads to the
picture that due to the coherent admixture of $f$-states to the
system, there will be an effective enhancement of $\Gamma(\omega)$
close to the Fermi energy. In this sense one may identify the physics in
this region of parameter space with the picture of coherent Kondo
scatterers.

The most interesting question is how the energy scales
\begin{figure}
\begin{center}\mbox{}
\psfig{figure=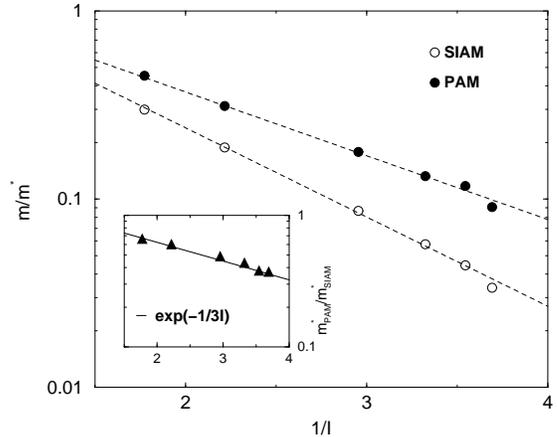,angle=-90,width=0.4\textwidth}\\
\end{center}
\caption[]{Inverse effective mass as function of $1/I$ (see text) for
  $\varepsilon_c=0$ and $\varepsilon_f=-1$. The inset shows 
  the ration $m^\ast_{\rm PAM}/m^\ast_{\rm SIAM}$ together with a fit
  $m^\ast_{\rm PAM}/m^\ast_{\rm SIAM}=e^{-1/3I}$.}
\label{fig:4}
\end{figure}
\noindent of the dilute
system and the lattice, $T_{\rm K}$ and $T_0$, are related in this
parameter regime. Let us recall that from the Gutzwiller ansatz one
obtains \cite{ueda} $T_{\rm
  K}/T_0=m^\ast_{\rm PAM}/m^\ast_{\rm SIAM}=\exp(-1/2I)$, where
$I=8\rho_0^c(0)V^2/U$; this prediction was found to be consistent
with recent DMFT QMC simulations \cite{PAMinfd2} where the Wilson
Kondo scale was estimated form the excess impurity susceptibility.

At $T=0$, the most efficient way to extract the 
low-energy scale is by calculating the effective mass according to
(\ref{equ:7}). With $m/m^\ast\propto T_0$ we are then able to discuss
the variation of $T_{\rm K}$ and $T_0$ with $U$. The result is shown
in Fig.\ \ref{fig:4}, where $m/m^\ast$ is plotted versus $1/I$ and
$I=2\rho_0^c(0)V^2U/\left(|\varepsilon_f|(\varepsilon_f+U)\right)$ is the
Schrieffer-Wolff exchange coupling. As already predicted from the
spectra and self energies, the general relation $T_0>T_{\rm K}$
holds. In addition, both $T_{\rm K}$ and
$T_0$ apparently behave exponentially as function in $1/I$, but with
different slopes; in qualitative agreement with the predictions the
slope of $T_0$ is smaller. To quantify the relation between $T_{\rm
  K}$ and $T_0$ further we show in the inset to Fig.\ \ref{fig:4} the
ratio $T_{\rm K}/T_0=m^\ast_{\rm PAM}/m^\ast_{\rm SIAM}$ as function of
$1/I$. Again, an exponential behavior is observed; however, in
contrast to the predicted factor $1/2$ we find $\ln(T_{\rm
  K}/T_0)=-1/3I$ from
\begin{figure}
\begin{center}\mbox{}
\psfig{figure=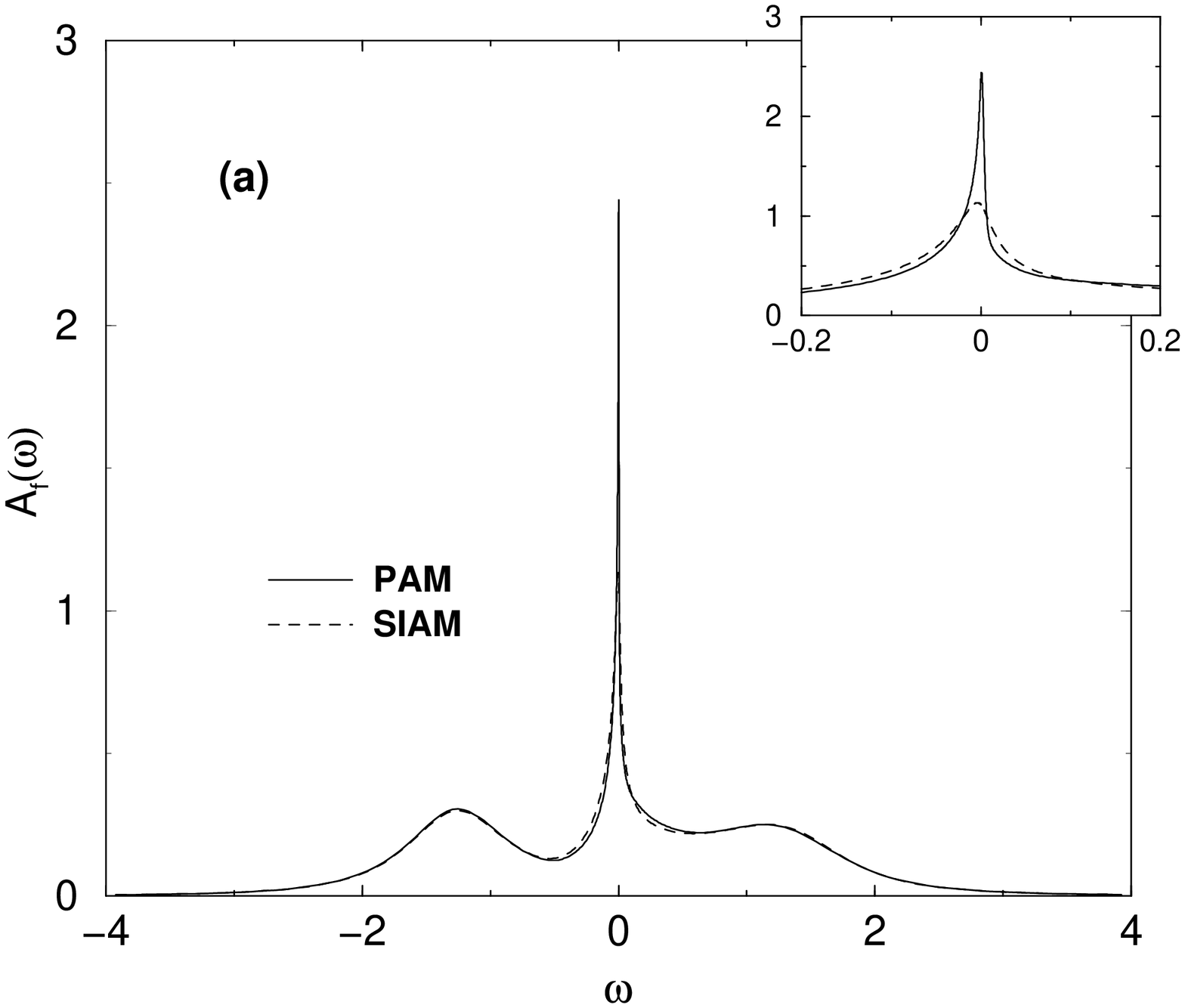,angle=-90,width=0.4\textwidth}\\
\mbox{}
\psfig{figure=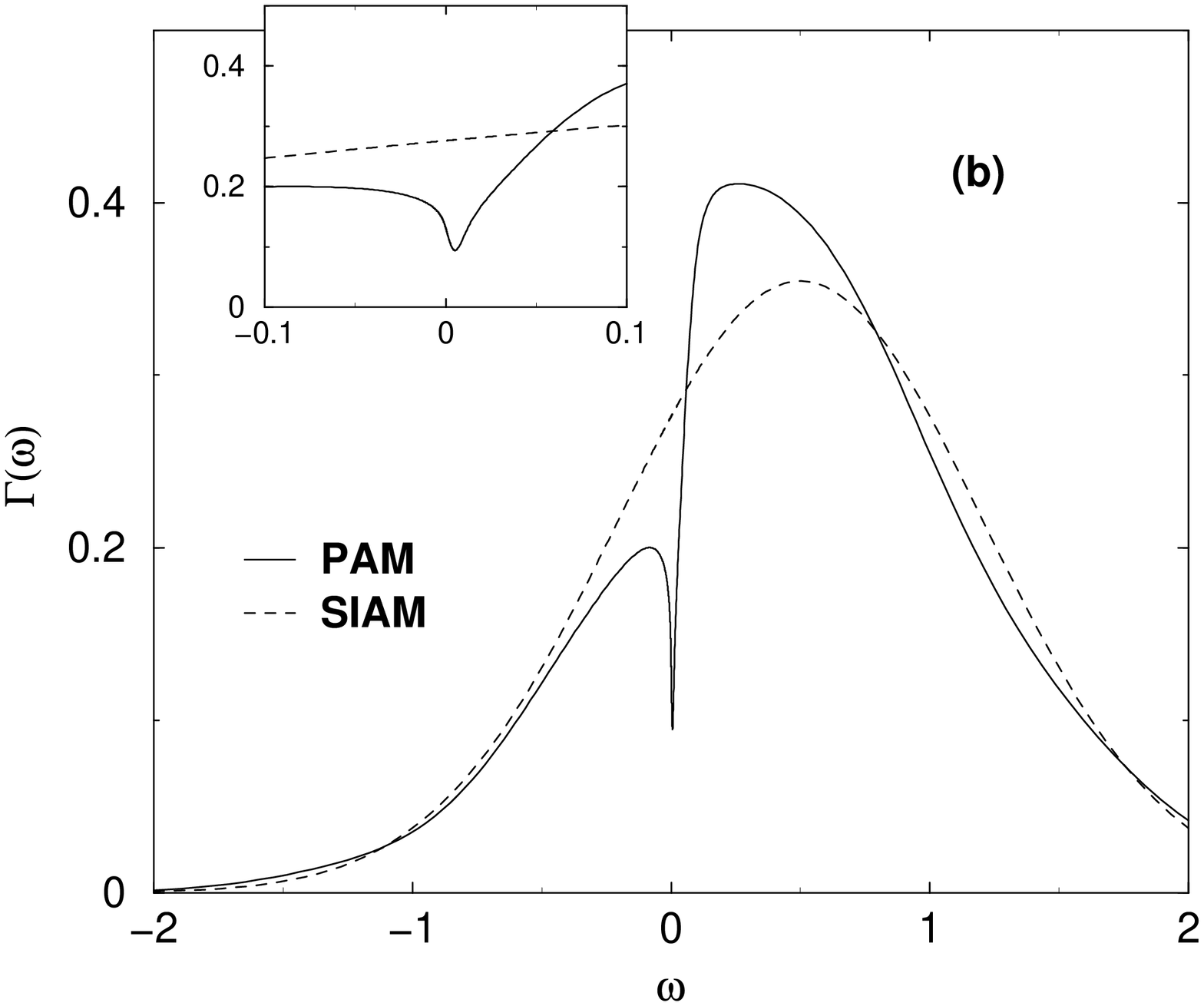,angle=-90,width=0.4\textwidth}
\end{center}
\caption[]{$f$-density of states $A_f(\omega)$ (a) and effective
  hybridization function $\Gamma(\omega)$ (b) for PAM (dashed line) and
SIAM (full line). The parameters are $\varepsilon_c=0.5$,
$2\varepsilon_f+U=0$, $U=2$ and $V^2=0.2$.\label{fig:5}}
\end{figure}
\noindent our data. For the time being we do not have an
explanation for this discrepancy between the QMC and NRG results;
whether it might be related to different schemes to extract $T_0$ --
scaling behavior of the excess susceptibility as function of
temperature in the QMC versus effective mass at $T=0$ in the NRG --
has to be clarified.

The second possibility to destroy particle-hole symmetry is to choose
$\varepsilon_c\ne0$. As a typical result for that parameter regime we
show in Fig.\ \ref{fig:5} the DOS $A_f(\omega)$ (Fig.\ \ref{fig:5}a)
and the corresponding effective hybridization function
$\Gamma(\omega)$ (Fig.\ \ref{fig:5}b) for $\varepsilon_c=0.5$,
$2\varepsilon_f+U=0$, $U=2$, and $V^2=0.2$. For the filling we obtain
$n_f\approx1$ and $n_c\approx0.6$ for
both the PAM (dashed line) and the single impurity Anderson model
(SIAM, full line). As usual, one sees the characteristic structures, namely the
charge-excitation peaks at $\omega\approx\pm U/2$  and the Kondo
resonance at the Fermi level. However, in contrast to the results with 
$\varepsilon_c=0$, we do not find any hint of a hybridization gap in
the lattice DOS; in fact, the DOS looks pretty much like that of a
conventional SIAM. The major difference to the DOS of the SIAM is an
enhancement of the ASR and a reduction of its width \cite{niki3}, as is apparent
from the inset to Fig.\ \ref{fig:5}a. Particularly interesting is the
behavior of the effective hybridization function in Fig.\
\ref{fig:5}b. Similar to the case $\varepsilon_c=0$,
$2\varepsilon_f+U>0$ it is considerably reduced in the region around the 
Fermi level; in contrast to the former case, however,
the sharp quasi-particle contribution is missing and
the average value of $\Gamma(\omega)$ over the region of the order $T_0$ around
$\omega=0$ is still reduced from the non-interacting value here.
The depletion in $\Gamma(\omega)$ around the Fermi level has been
coined as hall-mark of exhaustion physics in the PAM and related
models \cite{niki3,krish99}, since according to Nozi\`eres phenomenological
picture \cite{nozieres2} the effective density of medium states available 
at a given site should be reduced due to screening at other sites.

The \hfill fundamental  \hfill difference \hfill in \hfill the \hfill
physics \hfill between
\begin{figure}
\begin{center}\mbox{}
\psfig{figure=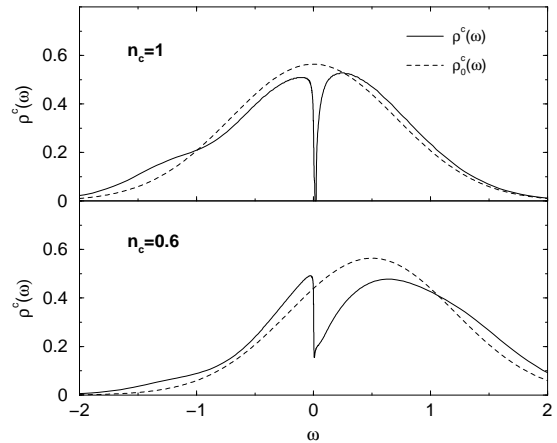,angle=-90,width=0.4\textwidth}
\end{center}
\caption[]{$c$-density of states $\rho^c(\omega)$ of the PAM 
for $\varepsilon_f=-1$, $V^2=0.2$ and $U=6$, $\varepsilon_c=0$
($n_c\approx1$, upper panel) and $U=2$, $\varepsilon_c=0.5$
($n_c\approx0.6$, lower panel). The dashed lines depict the bare
conduction DOS $\rho^c_0(\omega)$ for the corresponding $\varepsilon_c$.\label{fig:5a}}
\end{figure}
\noindent
$n_c\approx1$ and $n_c\ll1$ also manifests itself in the behavior of
the DOS of the conduction states. Typical results for this quantity
are shown in Fig.\ \ref{fig:5a} for $\varepsilon_f=-1$, $V^2=0.2$ and
$\varepsilon_c=0$, $U=6$ ($n_c\approx1$, upper panel) and
$\varepsilon_c=0.5$, $U=2$ ($n_c\approx0.6$, lower panel). For comparison,
the bare band DOS for the corresponding value of $\varepsilon_c$ is also
included (dashed curves);
the choice of different values of $U$ for the cases $\varepsilon_c=0$
and $\varepsilon_c=0.5$ is necessary to ensure that both systems are
metallic. One observes fundamental qualitative differences in the DOS, 
especially close to the Fermi surface, which in our opinion are
related to the different physical properties of the different regimes
and do not depend critically on a particular choice of the
interaction strength.
For $n_c\approx1$, the appearance of a gap slightly above the Fermi
level in the conduction DOS again supports the notion of
hybridizing bands in this region of parameter space.
On the other hand, the conduction DOS for the
case $n_c\approx0.6$ does not show the typical form of hybridized
bands, but merely a pseudo-gap at the Fermi energy. The fact, that the spectrum only develops a pseudo-gap with
finite DOS for $\omega=0$, can again be understood as sign of
exhaustion, since the formation of a full hybridization gap requires
complete Kondo screening by the band states to occur {\em for each f-site};
while the formation of a pseudo-gap can be interpreted that only part
of the f-sites are screened by the conduction electrons.

The reduced effective hybridization around the Fermi level observed in 
Fig.\ \ref{fig:5} gives
also rise to a reduced low-energy scale, characterized by an effective
mass  $m^\ast/m\approx17$  in the PAM, whereas the corresponding
quantity for the SIAM is $m^\ast/m\approx8$. The behavior of the
low-energy scale \hfill as \hfill function \hfill of \hfill $n_c$ \hfill for \hfill fixed \hfill $U=2$, \hfill $V^2=0.25$ \hfill and
\begin{figure}
\begin{center}\mbox{}
\psfig{figure=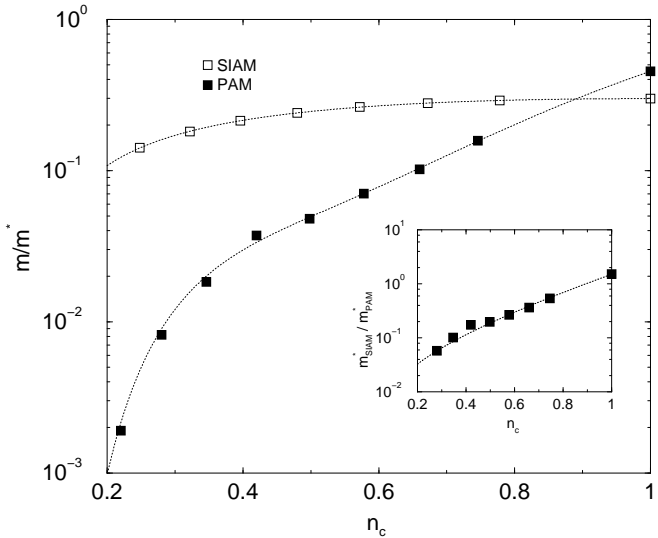,angle=-90,height=0.4\textwidth}
\end{center}
\caption[]{$m/m^\ast$ vs.\ $n_c$ for $U=2$, $V^2=0.25$ and $n_f\approx1$. The lines
are guides to the eye.\label{fig:6} The inset shows the ratio 
$m^\ast_{\rm SIAM}/m^\ast_{\rm PAM}\equiv T_0/T_{\rm K}$ versus $n_c$. 
The data are fit by $T_0/T_{\rm K}\propto n_c\exp\left(c\cdot n_c\right)$ 
(full line), where $c\approx5/2$.}
\end{figure}
\noindent
$n_f\approx1$ is shown in Fig.\ \ref{fig:6}.
Note that for $n_c\approx1$ the value $T_0$ is again
enhanced over the impurity scale with $\ln(T_{\rm K}/T_0)=-1/3I$.
Below $n_c\approx0.8$ the ratio $T_0/T_K$ decreases below one and falls
rapidly and monotonically with $n_c$, being almost two orders of magnitude
smaller for $n_c\approx0.2$.  The ratio 
$m^\ast_{\rm SIAM}/m^\ast_{\rm PAM}\equiv T_0/T_{\rm K}$ is shown in the
inset to Fig.\ \ref{fig:6}. This ratio falls more rapidly than $T_{\rm K}$, 
i.e.\ $T_0/T_{\rm K}^2$ is not constant. The ratio $T_0/T_{\rm K}$ can be fit to a form 
$T_0/T_{\rm K}\propto n_c\cdot\exp\left(c\cdot n_c\right)$ with $c\approx5/2$
(Full curve in the inset to Fig.\ \ref{fig:6}). 
It gives  an excellent account of the data. This behaviour especially means
that $T_0\sim n_c T_{\rm K}$ as $n_c\to0$ \cite{gabi_old}.
Note that the Gutzwiller result for $n_c\ll1$ predicts {\em both}\/ $T_0$ and
$T_{\rm K}$ to behave like $n_ce^{-c\cdot n_c}$, but (i) predicts $T_0/T_{\rm K}>1$
and (ii) clearly gives no proportionality to $n_c$ in the ratio $T_0/T_{\rm K}$.

Nozi\`eres phenomenological arguments also
lead to an estimate of $T_0$ as function of $n_c$,
namely $T_0\propto (T_K)^2/\rho_0^c(0)$ \cite{nozieres2}. This relation has
recently been tested with IPT \cite{krish99} and,
within this
approach, found to be fulfilled
at least for $U/V^2\approx 4$ between $0.4\le n_c\le0.8$ when $T_0$ is
estimated from the width of the Kondo resonance. We first note,
that clearly our result in Fig.\ \ref{fig:6} is not compatible with this
prediction. In order to clarify the relation \hfill between \hfill $T_0$ \hfill and \hfill $T_{\rm K}$
\hfill we \hfill compare
\hfill $T_0$ \hfill and \hfill $T_K$
\begin{figure}
\begin{center}\mbox{}
\psfig{figure=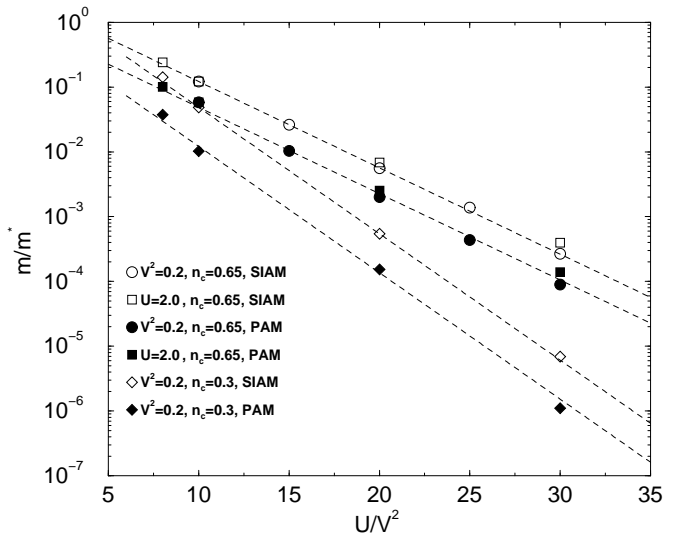,angle=-90,height=0.4\textwidth}
\end{center}
\caption[]{$m/m^\ast$ as function of $U/V^2$ for $n_c\approx0.6$
  (squares and circles) and $n_c\approx0.31$ (diamonds) with $n_f\approx1$.
The circles (squares) denote varying $U$ ($V^2$) for fixed $V^2=0.2$
($U=2$).
The diamonds show the behaviour of $m/m^\ast$ for a different filling
$n_c\approx0.3$ and $V^2=0.2$ for varying $U$.\label{fig:7}}
\end{figure}
\noindent as function of $U/V^2$ for fixed $n_c$.
The results for $n_c\approx0.6$ as
function of $U/V^2$ for both varying $U$ at constant $V^2=0.2$
(circles) and varying $V^2$ at constant $U=2$ (squares) is shown in
Fig.\ \ref{fig:7} on a semi-logarithmic scale; to study the dependence
on the filling $n_c$, we included calculations
for $n_c=0.31$ (diamonds, varying $U$ only). Evidently
$T_K$ and $T_0$ both follow an exponential law $T_0,\;
T_K\propto\exp(-a/I)$, where $I=8\rho^c_0(0)V^2/U$ is the
Schrieffer-Wollf exchange coupling for $2\varepsilon_f+U=0$.
However, the curves for the SIAM and PAM for fixed $n_c$ are parallel
in the semi-logarithmic plot in Fig.\ \ref{fig:7}, i.e.\ the
coefficients of $U/V^2$ in the exponents of both quantities are identical. 
This of course means $T_0\propto T_K$ rather than $T_0\propto (T_K)^2$, as
predicted by Nozi\`eres. Thus, neither as function of $n_c$ nor as function
of $U/V^2$ does the lattice scale $T_0$ obey Nozi\`eres prediction.

\section{Summary and Conclusion}\label{sec:5}
We have presented results for the PAM obtained within the DMFT at
$T=0$ using Wilson's NRG approach.
For the range of parameters studied 
here, the system can always be
characterized as a Fermi liquid with a strongly enhanced effective mass;
this lattice scale $T_0$ is enhanced over a corresponding impurity
Kondo scale $T_{\rm K}$ for the particle-hole symmetric conduction band in
accordance with perturbational results \cite{kuramoto,grewe,NCA2} or those
from Gutzwiller ansatz \cite{ueda}. Moreover, the picture of hybridizing
quasi particle bands leading to (pseudo-) gaps in the DOS was found to be 
valid here. On the other hand, in the case of an asymmetric conduction 
band and especially for low carrier concentration $n_c\alt0.8$, the 
spectral functions and corresponding effective hybridization functions 
show the signs of exhaustion and we observe a corrresponding strong
reduction of the lattice scale $T_0$ \cite{niki1,nozieres2}.  

        Together with our results, an extensive picture of exhaustion 
physics in the infinite-dimensional PAM has begun to emerge.  Close to half 
filling the low-temperature properties of the model can be characterized by 
one energy scale $T_0>T_K$; whereas away from half filling, two scales are 
apparent: $T_K$, where screening begins, and $T_0\ll T_K$ where coherence 
sets in.  At low temperatures $T\approx T_0$, the quasi particle features in 
the single-particle spectra become pronounced and have predominantly f
character \cite{niki2}.  Since only the states near the Fermi energy can 
participate in screening, the f-electron moments are screened predominantly 
by other f-electron states.  At a temperature $T_m$ ($T_0 < T_m < T_K$), a 
dip begins to develop in the effective hybridization rate at the Fermi 
surface $\Gamma(\omega\approx 0)$ \cite{niki3}, indicating that the states 
available for screening are becoming exhausted. This is a direct confirmation 
of the qualitative features of Nozi\`eres exhaustion scenario.  Nevertheless, 
for fixed $n_c$, we find that $T_0\propto T_K$, and as the conduction band 
filling changes we find $T_0\propto n_c \exp(c n_c)$, both are in direct 
contradiction with the predictions of Nozi\`eres.  Thus, we conclude that 
Nozi\`eres phenomenological arguments are too crude to capture the 
quantitative features of exhaustion.

To our knowledge there has not yet been a systematic experimental study 
concentrating on the signatures of exhaustion in metallic HF compounds.  
However, there are several experimentally relevant consequences of 
exhaustion predicted by DMF calculations.  Most of these predictions 
are related to the presence of two relevant scales and the protracted 
behavior of measurements as a function of temperature in crossover 
regime between these scales \cite{niki1}.  For example, as compared to 
the predictions of the SIAM with the same $T_0$, the photoemission peak 
should evolve much more slowly with temperature \cite{niki2}.  In 
addition, it should have significantly more weight since the height of the 
peak goes like $1/\Gamma(0)$ and its width is set by $T_0$.  Although these 
features have been reported in photoemission experiments on Yb-based
HF materials \cite{jjaa}, these results remain controversial \cite{debate} 
and it has been suggested that the experimental spectrum is representative
of the surface, and not the bulk\cite{allen_ybsurface}.  Fortunately, 
transport and neutron scattering experiments probe much further into the
bulk, and should display characteristic features due to exhaustion.  The 
calculated resistivity 
displays a non-universal peak and two other regions typical of HF
systems and associated with impurity-like physics at high temperatures 
$T\approx T_K$ and Fermi liquid formation at low $T \alt T_0$ \cite{niki3}.  
The peak resistivity shows features characteristic of exhaustion.  It occurs 
at $T\approx T_m$, the temperature at which exhaustion first becomes apparent 
as a dip $\Gamma(\omega\approx 0)$.  $T_m$ is non-universal in that it 
{\em{increases}} with decreasing $n_c$, $T_K$ and $T_0$. The Drude 
peak in the optical conductivity persists up to much higher temperatures
than predicted by the SIAM and the Drude weight weight rises dramatically 
with temperature\cite{niki3}.
The quasielastic peak in the angle integrated dynamic spin susceptiblity
also evolves more slowly with temperature than predicted by the SIAM,
and it displays more pronounced charge-transfer features \cite{us}.

        Despite the rich picture which has begun to emerge from DMF 
calculations, many questions remain unresolved.  Among these, three seem 
most prominent to us.  First, it is unclear what is happening as $n_c\to 1$.  
In this regime, following Nozi\`eres argument, there should also remain too
few states to screen the moments so the exhaustion scenerio would seem
to apply, too; nevertheless, the $T_0$ is enhanced relative to $T_K$
and all effects of exhaustion vanish.
It is tempting to attribute this vanishing of exhaustion to another
energy scale associated with the bare gap that appears in the spectra
as $n_c\to1$.
However, in recent calculations for a model with f-d hopping such that the 
hybridization $V_k\propto \epsilon_k$, where there is no conduction band
gap when $n_f=n_c=1$, in the regime of strong f-d hybridization $T_0\gg T_K$ 
is again found, suggesting that there must be some more fundamental 
reason for the absence of exhaustion \cite{king}.
Thus, these latter results together with the ones presented here,
surely demand for a critical reinvestigation of the phenomenology of
exhaustion.

Second, thus far, all calculations are for the orbitally
non-degenerate models. The effect of orbital degeneracy and crystal
field splittings has yet to be determined. However, in the limit of
infinite orbital degeneracy, the Kondo scale would seem to be
unrenormalized.

Third, following Doniach's arguments, RKKY interaction and
superexchange will compete with Kondo screening in the formation of
the ground state. In the present work, we have explicitely
concentrated on the paramagnetic state, i.e.\ these types of exchange
do not enter the calculation. However, within the DMF it is possible to 
study the influence of RKKY or superexchange on the mean-field level
by either looking at the susceptibility \cite{PAMinfd2} or allowing for a
symmetry-broken state. Generally, since the RKKY exchange grows like
$J^2$, and the Kondo scale is exponential in $-1/J$, Kondo screening
and hence exhaustion is expected to be most pronounced when the Kondo
exchange $J\sim V^2/U$ is large. However, we have found that
exhaustion can dramatically reduce the relevant low-energy scale;
which may extend the region where the magnetic exchange dominates in
the formation of the ground state. Thus a systematic study of the
magnetic phase diagram as function of $U/V^2$ is clearly desirable.

\noindent{\bf Acknowledgements:} It is a pleasure to acknowledge fruitful
discussions with
D.~Logan,
H.R.~Krishanmurthy,
F.~Anders,
A.~Georges,
N.~Grewe,
J.~Keller,
D.~Vollhardt. 
One of us (MJ) would like to acknowledge support by the NSF via grants
DMR9357199 and DMR 9704021. This work was in part supported by a NATO
collaborative research travel grant.


\begin{thebibliography}{99}
\bibitem{self_consistency} For a detailed review of the DMFT and the
  self-consistency cycle we refer the reader to \cite{infdrev,Geo96}.
\end{thebibliography}
\end{document}